\documentclass[seceqn]{elsart}

\usepackage{color}
\usepackage{framed}
\usepackage{slashed}
\usepackage{graphicx}

\usepackage{feynmp}

\usepackage{amsfonts}
\usepackage{amsmath}
\usepackage{mathrsfs}

\newcounter{bla}

\newcommand{\sherpa}{{\sc Sherpa}}

\newcommand{\gosam}{{\sc GoSam}}
\newcommand{\citegolem}{{%
\cite{Mastrolia:2010nb,Heinrich:2010ax,Cullen:2011ac}}}
\newcommand{\mgme}{{\sc MadGraph/MadEvent}}
\newcommand{\herwig}{{\sc Herwig}}
\newcommand{\alpgen}{{\sc Alpgen}}
\newcommand{\helac}{{\sc Helac}}
\newcommand{\comix}{{\sc Comix}}
\newcommand{\madgraph}{{\sc MadGraph}}

\newcommand{\feynrules}{{\sc FeynRules}}
\newcommand{\mathematica}{{\sc Mathematica}}
\newcommand{\feynarts}{{\sc FeynArts}}
\newcommand{\formcalc}{{\sc FormCalc}}
\newcommand{\whizard}{{\sc Whizard}}
\newcommand{\calchep}{{\sc CalcHep}}
\newcommand{\lanhep}{{\sc LanHep}}
\newcommand{\comphep}{{\sc CompHep}}

 \definecolor{lightgrey}{gray}{0.9}

 \def\btab#1\etab{\begin{tabular}{p{50mm}p{70mm}}#1\end{tabular}}
\def\btabx#1\etabx{\begin{tabular}{p{65mm}p{55mm}}#1\end{tabular}}
\def\btaby#1\etaby{\begin{tabular}{p{20mm}p{100mm}}#1\end{tabular}}
\def\btabwide#1\etabwide{\begin{tabular}{p{82mm}p{38mm}}#1\end{tabular}}
 \def\bcen{\begin{center}}
 \def\ecen{\end{center}}
 
\def\bgfb#1\egfb{\bcen\fcolorbox{black}{lightgrey}{\parbox{130mm}{\btab#1\etab}}\ecen}

\def\bgfbx#1\egfbx{\bcen\fcolorbox{black}{lightgrey}{\parbox{130mm}{\btabx#1\etabx}}\ecen}

\def\bgfbalign#1\egfbalign{\bcen\fcolorbox{black}{lightgrey}{\parbox{130mm}{\btaby#1\etaby}}\ecen}
\def\bgfbwide#1\egfbwide{\bcen\fcolorbox{black}{lightgrey}{\parbox{130mm}{\btabwide#1\etabwide}}\ecen}

\newcommand{\eg}[0]{{\it e.g.}}
\newcommand{\ie}[0]{{\it i.e.}}

\newcommand{\comment}[1]{}

\newcommand{\python}{{\sc Python}}

\begin{document}

%%%%%%%%%%%%%%%%%%%%%%%%%%%%%%%%%%%%%%%%%%%%%%%%%%%%%%%%%%%%%%%%%%%%%%%%
% Shamelessly stolen from Thorsten's thohacks.sty
%%%%%%%%%%%%%%%%%%%%%%%%%%%%%%%%%%%%%%%%%%%%%%%%%%%%%%%%%%%%%%%%%%%%%%%%
\catcode`\@=11
\font\manfnt=manfnt
\def\Watchout{\@ifnextchar [{\W@tchout}{\W@tchout[1]}}
\def\W@tchout[#1]{{\manfnt\@tempcnta#1\relax%
  \@whilenum\@tempcnta>\z@\do{%
    \char"7F\hskip 0.3em\advance\@tempcnta\m@ne}}}
\let\foo\W@tchout
\def\dubious{\@ifnextchar[{\@dubious}{\@dubious[1]}}
\let\enddubious\endlist
\def\@dubious[#1]{%
  \setbox\@tempboxa\hbox{\@W@tchout#1}
  \@tempdima\wd\@tempboxa
  \list{}{\leftmargin\@tempdima}\item[\hbox to 0pt{\hss\@W@tchout#1}]}
\def\@W@tchout#1{\W@tchout[#1]}
\catcode`\@=12
%%%%%%%%%%%%%%%%%%%%%%%%%%%%%%%%%%%%%%%%%%%%%%%%%%%%%%%%%%%%%%%%%%%%%%%%

\begin{frontmatter}

\begin{flushright}
CP3-11-25,
IPHC-PHENO-11-04,
IPPP/11/39,
DCPT/11/78,
MPP-2011-68
\end{flushright}

\title{UFO - The Universal FeynRules Output}

\author[a]{C\'eline Degrande},
\author[b]{Claude Duhr},
\author[c]{Benjamin Fuks},
\author[b]{David Grellscheid},
\author[a]{Olivier Mattelaer},
\author[d]{Thomas Reiter}

\address[a]{Universit\'e catholique de Louvain,\\ Center for particle physics and phenomenology (CP3),\\ Chemin du cyclotron, 2, B-1348 Louvain-La-Neuve, Belgium\\ Email: celine.degrande@uclouvain.be, olivier.mattelaer@uclouvain.be}
\address[b]{Institute for Particle Physics Phenomenology, University of Durham,\\ Durham, DH1 3LE, United Kingdom,\\ Email: claude.duhr@durham.ac.uk, david.grellscheid@durham.ac.uk}
\address[c]{Institut Pluridisciplinaire Hubert Curien/D\'epartement Recherches
     Subatomiques, Universit\'e de Strasbourg/CNRS-IN2P3, 23 Rue du Loess, F-67037
     Strasbourg, France\\ E-mail address: benjamin.fuks@iphc.cnrs.fr}
\address[d]{Max-Planck-Institut f\"ur Physik,\\
F\"ohringer Ring~6, 80805~M\"unchen, Germany\\
Email: reiterth@mpp.mpg.de}

\begin{abstract}
We present a new model format for automatized matrix-element generators, 
the so-called \textit{Universal FeynRules Output (UFO)}. The format
is universal in the sense that it features compatibility with
more than one single generator and is designed to be flexible, modular
and agnostic of any assumption such as the number of particles or the color and Lorentz structures appearing in the interaction
vertices. Unlike other model formats where text files need to be parsed, 
the information on the model is encoded into a {\sc Python}\ module that can easily be 
linked to other computer codes. We then describe an
interface for the \mathematica\ package \feynrules\ that allows for an
automatic output of models in the UFO format.


\end{abstract}

\end{frontmatter}

\newpage

\section{Introduction}
Monte Carlo simulations of the physics to be observed at the Large
Hadron Collider (LHC) at CERN play a central role in the exploration of the electroweak
scale, both from the experimental point of view of establishing an
excesses over the expected Standard Model (SM) backgrounds as well as from the
phenomenological point of view by providing possible explanations for the observations.
For this reason, activities in the field of Monte Carlo simulations have been
rather intense over
the last fifteen years, resulting in many advances in the field. 
Automated tree-level matrix-element generators, such as \alpgen\ \cite{Mangano:2002ea},
\comix\ \cite{Gleisberg:2008fv}, \comphep/\calchep\
\cite{Pukhov:1999gg,Boos:2004kh,Pukhov:2004ca}, \helac\ \cite{Cafarella:2007pc},
\herwig\ \cite{Corcella:2000bw,Bahr:2008pv},
\mgme\ \cite{Stelzer:1994ta,Maltoni:2002qb,Alwall:2007st,Alwall:2008pm,Alwall:2011uj},
\sherpa\ \cite{Gleisberg:2003xi,Gleisberg:2008ta}, or \whizard\
\cite{Moretti:2001zz,Kilian:2007gr}, describing the hard scattering processes
where the Beyond the Standard Model (BSM) physics is expected to show up have
been developed. As a consequence, the problem of the automatic generation of
tree-level matrix elements for a large class of Lagrangian-based BSM theories
is solved, at least in principle. 

Due to the numerous existing BSM theories based on ideas in constant
evolution, the
implementation of these models into Monte Carlo event generators remains a
tedious and error-prone task. Feynman rules associated with a given BSM model
must be derived and then implemented one at the time into the various codes,
which often follow their own specific conventions and formats. A first step in
the direction of automating this procedure by starting directly from the
Lagrangian of the model has been
made in the context of the \lanhep\ package \cite{Semenov:1998eb} linked to the
\comphep\ and \calchep\ programs. Recently, a new efficient framework going beyond
this scheme has been developed. It is based on the \feynrules\ package
\cite{Christensen:2008py,Christensen:2009jx,Christensen:2010wz,Duhr:2011se} and
proposes a general and flexible environment allowing to develop a model,
investigate its phenomenology and eventually confront it to data. Its
virtue has been illustrated in the context of the \comphep/\calchep,
\feynarts/\formcalc\ \cite{Hahn:2000kx}, \mgme, \sherpa\ and \whizard\ programs,
by implementing several new physics theories in \feynrules\ and then passing
them to the different tools for a systematic validation procedure. The
approach is based on a modular structure where each node consists in an
interface to a dedicated matrix-element generator. Since the latter have in
general hard-coded information regarding the supported Lorentz and/or color structures, the
interfaces check whether a given vertex is compliant with a given matrix-element generator, in which case the vertex is written to file in a format suitable for the generator. The final output consists then in a set of
text files that can be used in a similar way to any other built-in model.

The procedure spelled out above, where communication between
\feynrules\ and the matrix-element generators proceeds exclusively via a set of
well-defined text files that must be parsed and interpreted, has some serious
limitations. In particular, extending the format to include more general
structures, like higher-dimensional operators and/or non-standard color
structures, is difficult to incorporate into a static text-based format. In this
paper we present a new format, dubbed the Universal \feynrules\ Output or the
UFO, for model files that goes beyond existing formats in various ways. The
format is completely generic and, unlike existing formats, it does not make any
\emph{a priori} assumptions on the structures that can appear in a model. The
aim is to provide a flexible format, where \emph{all} the information about a
model is represented in an abstract form that can easily be accessed by other
tools. The information on the
particles,
parameters and vertices of the model are stored in a set of \python\ objects, each of them
being associated with a list of
attributes related to their properties. 
%% REVISED
This way of representing the model information has some benefits
over the more traditional plain text table-based format, because it
allows, \eg, to add a missing piece of information directly as a new attribute
to an existing object. As an example, extending a table-based format to accommodate 
higher-point vertices requires to change the format of the table and to adapt the readers
for the table accordingly. In an object-oriented format like the UFO, the same extension is 
trivial, as the number of particles entering a vertex is just an attribute of the vertex, 
so no extension of rewriting of the readers is necessary.
%flexible and extendable since, \eg, a missing piece of information can be
%added as a new attribute to an existing object.
%The format has the benefit to be fully
%flexible and extendable since, \eg, a missing piece of information can be
%added as a new attribute to an existing object. 
Presently, the UFO format is already 
used by the \madgraph\ version 5 \cite{Alwall:2011uj}  and the \gosam\ 
generators~\citegolem, and will be used in a near future by \herwig++.
%\and \golem\ 2{.}0~\citegolem.

The paper is organized as follows. In Section \ref{sec:ufo_format}, we describe
the features of the UFO format as a stand-alone \python\ module, while
Section \ref{sec:feynrules} addresses the automation of  writing an UFO
model through \feynrules. Section \ref{sec:nlo} is dedicated to the UFO features
beyond tree level and in Section \ref{sec:example}, we provide an example
of how to implement a model containing non-trivial
Lorentz structures with the help of \feynrules\ into \madgraph\ 5.
Our conclusions are drawn in Section \ref{sec:conclusions}.

\section{The UFO format}
\label{sec:ufo_format}
Any quantum field theory can be defined by a threefold information,
\begin{itemize}
\item a set of particles, defined together with their quantum numbers (spin,
electric charge, \emph{etc.}),
\item a set of parameters (masses, coupling constants, \emph{etc, ...}),
\item a Lagrangian describing the interactions among the different particle species.
\end{itemize}

However, matrix-element generators do not work, in general, directly with the Lagrangian,
but rather with an explicit set of vertices. In the rest of this section, we assume that we have 
extracted all the vertices from the Lagrangian of
a given model and only restrict ourselves to describing a new generic
format to implement the information on the
particles and parameters of the model along with the vertices describing the
interactions among the particles into matrix-element generators. 
The issue of the extraction of the vertices from
the Lagrangian and their translation into this new format in an automated
fashion via the \feynrules\ package will be discussed in
Section~\ref{sec:feynrules}.

The Universal \feynrules\ Output (UFO) allows to translate all the information
about a given particle physics model into a \python\ module that can
easily be linked to existing matrix-element generators. While in general each 
generator is following its own format and conventions, the UFO format
goes beyond this approach in the sense that it is, by definition, not tied to any
specific matrix-element generator. More specifically, it saves the model information in an abstract
(generator-independent) way in terms of \python\ objects. An UFO
model is hence a standalone \python\ module, containing ready-to-go
definitions for all the classes representing particles, parameters, \emph{etc.},
and which can be directly linked to an existing
matrix-element generator without any modification or further interfacing. 

In this section we give a detailed account on the UFO format, putting
special emphasis on the
definition of the different classes useful for designing model files. In
general, an UFO model consists of a directory containing a set of text files
that can be split into two distinct classes,
\begin{itemize}
\item Model-independent files:
\begin{itemize}
\item[-] \verb+__init__.py+,
\item[-] \verb+object_library.py+,
\item[-] \verb+function_library.py+,
\item[-] \verb+write_param_card.py+,
\end{itemize}
\item Model-dependent files:
\begin{itemize}
\item[-] \verb+particles.py+,
\item[-] \verb+parameters.py+,
\item[-] \verb+vertices.py+,
\item[-] \verb+couplings.py+,
\item[-] \verb+lorentz.py+,
\item[-] \verb+coupling_orders.py+.
\end{itemize}
\end{itemize}
Since the UFO format is based on the \python\ language, all files have a 
{\tt .py} extension. The model-independent files are identical for every
model and contain, among others, the definitions of the classes which the
model-dependent objects (particles, parameters, {\emph{etc}.}) are instances
of. All those files are provided as self-contained \python\ modules.

%%%%%%%%%%%%%%%%%%%%%%%%%
%%%%%%%%%%%%%%%%%%%%%%%%%
%%%%%%%%%%%%%%%%%%%%%%%%%
%%%%%%%%%%%%%%%%%%%%%%%%%

\subsection{Initialization and structure of the objects and functions}

A file named \verb+__init__.py+ inside a directory is standard in the \python\
language and corresponds to a tag for importing complete
\python\ modules by issuing the \python\ command
\begin{verbatim}
import Directory_Name
\end{verbatim}
where \verb+Directory_Name+ refers to the name of the directory containing the
\verb+__init__.py+ file. However, in addition to the possiblity of importing a
complete UFO model, this file also contains, in the UFO case, links to
different lists of quantities associated with the various objects defined in a
model,
\begin{itemize}
\item  \verb+all_particles+
\item  \verb+all_vertices+ 
\item  \verb+all_couplings+
\item  \verb+all_lorentz+
\item  \verb+all_parameters+
\item  \verb+all_coupling_orders+
\item  \verb+all_functions+ 
\end{itemize}
These lists allow, \eg, to access the full particle content of a
model in an easy way downstream in the code. Moreover, every time that an instance
of a class is created in the model, it will be automatically added to the
corresponding list.

An UFO model can be fully implemented with the help of a small number of basic
classes, denoted {\tt Particle}, {\tt
Parameter}, {\tt Vertex}, {\tt Coupling}, {\tt Lorentz} and {\tt CouplingOrder}.
All of these classes are derived from the mother class {\tt
UFOBaseClass}, defining  a set of common methods and attributes accessible
in the same way by each class. The mother class, together with all its children, 
is defined in the file {\tt object\textunderscore library.py}. 
In particular, each class has methods to display all the attributes associated
to a given instance of the class,
as well as to return or set the values of these attributes. As an example, if {\tt P} is an instance of the 
class {\tt Particle} and if {\tt charge} is an attribute of this class (see Section \ref{subsec:particles.py}), the
charge of the corresponding particle can be accessed in the standard way by
issuing the command {\tt P.charge}. The complete list of attributes of the {\tt
UFOBaseClass} class is summarized in Table \ref{tab:default_python_commands}.

\begin{table}
\bgfbalign
\multicolumn{2}{c}{\textbf{Table~\ref{tab:default_python_commands}: Attributes and methods available to all UFO classes}}\\
\\
{\tt get\textunderscore all} & Returns a list of all the attributes of an object.\\
{\tt nice\textunderscore string} & Returns a string with a representation of an object containing the values associated with each of its attributes. \\
{\tt get} &  This method provides access to the value of an attribute of
an object. 
As an example, if {\tt P} denotes an instance of a class with attribute
{\tt charge}, then {\tt P.get('charge')} and
 {\tt P.charge} are equivalent means of accessing the value of the attribute
 {\tt charge}.\\ 
{\tt set}  & This method allows one to  modify the value of an attribute of an object. 
As an example, if {\tt P} denotes an instance of class with attribute {\tt
charge}, then {\tt P.set('charge', 1)}, or equivalently {\tt
P.charge = 1}, will set the attribute denoted by {\tt charge} to unity.
\egfbalign
\textcolor{white}{\caption{\label{tab:default_python_commands}}}
\end{table}

The file {\tt function\textunderscore library.py} is related to the
implementation of user-defined functions into an UFO module via the special
class {\tt Function}, which translates functions that can be defined within a
single \python\ line ({\it i.e.}, the so-called \python\ `lambda' functions) to
other programming languages (such as {\sc Fortran} or {\sc C++}). Let us note
that this specific type of
functions is currently the only type of user-defined functions supported by the
UFO format. A member of the class {\tt Function} contains three mandatory
attributes, called {\tt name}, {\tt arguments} and {\tt expression}. While {\tt
name} is a string representing the name of the function, the attributes {\tt arguments} and 
{\tt expression} correspond to a sequence of strings for the names of
the variables the function depends upon and a string representing the valid \python\ 
expression defining the function itself.
Several functions are by default included into the
UFO function library, 
\begin{itemize}
\item \verb+complexconjugate+: complex conjugation,
\item \verb+csc+: the trigonometric function cosecant,
\item \verb+acsc+: the cyclometric function arccosecant,  
\item \verb+im+: the imaginary part of a complex number,   
\item \verb+re+: the real part of a complex number,  
\item \verb+sec+: the trigonometric function secant,
\item \verb+asec+: the cyclometric function arcsecant.
\end{itemize}
These functions consist in a set of common mathematical functions for which the standard \python\
module {\tt cmath} is insufficient or unpractical. As an example, the cosecant
function {\tt csc} (not included in the {\tt cmath} library) is
implemented within the UFO module as an instance of the aforementioned class
{\tt Function} via
\begin{verbatim}
csc = Function(name = 'csc',
             arguments = ('z',),
             expression = '1./cmath.sin(z)')
\end{verbatim}

%%%%%%%%%%%%%%%%%%%%%%%%%
%%%%%%%%%%%%%%%%%%%%%%%%%
%%%%%%%%%%%%%%%%%%%%%%%%%
%%%%%%%%%%%%%%%%%%%%%%%%%
\subsection{Implementation of the particle content of a model} 
\label{subsec:particles.py}

In the UFO format, all particles are instances of the class {\tt Particle}
defined in the file
{\tt particles.py}. Even if the Lagrangian of a model is in general more easily
written in terms gauge eigenstates, matrix-element
generators usually work at the level of mass eigenstates. Hence only mass
eigenstates should be defined in the {\tt particles.py} file.

The definition of a particle might read, for, \eg, a top quark, as
\begin{verbatim}
  t = Particle( pdg_code = 6,
                name = 't',
                antiname = 't~',
                spin = 2,
                color = 3,
                mass = Param.MT,
                width = Param.WT,
                texname = 't',
                antitexname = '\\bar{t}',
                charge = 2/3,
                line = 'straight',
                LeptonNumber = 0
                )   
\end{verbatim}
The class {\tt Particle} has various attributes that are
summarized in Table \ref{tab:particle_attributes}. In the following we content ourselves
to highlight only the most important points. First, note that, apart from a set of mandatory arguments 
(all attributes but the last two in the example above),
the {\tt Particle} class can be given an arbitrary number of optional attributes
(the {\tt line} and {\tt
LeptonNumber} attributes
in the example).
There are three predefined optional attributes, which are summarized in Table \ref{tab:particle_attributes}.  Every additional
optional attribute must be an integer representing additional model-dependent additive quantum numbers 
(as the attribute {\tt LeptonNumber} in the example). The
only exceptions regarding the treatment of the quantum numbers concern the
electric charge and color representation, which are always mandatory and stored
in the attributes {\tt charge} and {\tt color}. 

A particle object is identified through its name, a string stored
in the {\tt name} attribute. In a similar
fashion, the attribute {\tt antiname} is a string representing the name of the corresponding antiparticle.
Note that self-conjugate particles,
\ie, particles that are their own antiparticles, are identified by having identical {\tt name} and {\tt antiname}
attributes (\emph{i.e.}, even for self-conjugate particles, the {\tt antiname} attribute must be defined). 
The transformation properties of the particle under the Lorentz group and the
QCD and electromagnetic gauge groups 
are specified through the {\tt spin}, {\tt color} and {\tt charge} attributes. Each of these attributes takes an integer value:
\begin{itemize}
\item {\tt spin}: the possible values are $2s+1$, where $s$ is the spin of
the particle. For the moment only $s\le 2$ is supported. By convention, the value $-1$ is used for ghost fields.
\item {\tt color}: the possible values are $1$, $\pm 3$, $\pm 6$ and $8$,
corresponding to singlets, (anti)triplets, (anti)sextets and octets. 
\item {\tt charge}: any rational number, representing the electric charge of the particle.
\end{itemize}
Inside matrix-element generators, particles are often identified through an integer number referring to
the Particle Data Group (PDG) numbering scheme \cite{Nakamura:2010zzi}. This
code is stored in the {\tt pdg\textunderscore code} attribute, which can be set
to any integer value, even though it is highly recommended to follow the
existing conventions whenever possible. Finally, masses and
widths are encoded in the {\tt mass} and {\tt width} attributes.
They refer to the corresponding instances of the {\tt Parameter} class defined 
in the file {\tt parameters.py} (see Section 
\ref{sec:parameters.py}). Therefore, at the beginning of the {\tt particles.py}
file, the {\tt Parameter} objects are imported via the \python\ instruction
\begin{verbatim}
  import parameters as Param
\end{verbatim} 
 
\begin{table}
\bgfbalign
\multicolumn{2}{c}{\textbf{Table~\ref{tab:particle_attributes}: Attributes of
the particle class}}\\
\\
{\tt pdg\textunderscore code} & An integer corresponding the identification
number related to the PDG numbering scheme \cite{Nakamura:2010zzi}. \\
{\tt name} &  A string specifying the name of the particle. \\
{\tt antiname} &  A string specifying the name of the antiparticle. If the
particle is self-conjugate, {\tt antiname} must be identical to {\tt name}.\\
{\tt spin}  & An integer corresponding to the spin of the particle in the form
$2s+1$. By convention, the spin of a ghost field (anticommuting scalar field) is -1.\\
{\tt color}  & An integer corresponding to the dimension of the color representation of
the particle ($1, \pm 3, \pm 6, 8$).\\
{\tt mass} & A {\tt Parameter} object corresponding to the mass of the
particle. If the particle is massless, the value must be set to {\tt Param.ZERO}.\\
{\tt width} & A {\tt Parameter} object corresponding to the width of the
particle. If the width is zero, the value must be set to {\tt Param.ZERO}.\\
{\tt texname} & A \TeX{} string representing the particle name.\\
{\tt antitexname} & A \TeX{} string representing the antiparticle name.\\
{\tt charge} & A rational number equal to the electric charge of the particle.\\
\multicolumn{2}{l}{\textbf{Optional attributes}}\\
{\tt goldstone} & A boolean, tagging a scalar field as a Goldstone boson ({\tt
true}) or not ({\tt false}). The default value is {\tt false}.\\
{\tt propagating} & A boolean, tagging the corresponding particle as auxiliary
and non-propagating ({\tt false}) field or as a physical field ({\tt true}). The
default value is {\tt true}.\\
{\tt line} & A string representing how the propagator of the particle should be
drawn in a Feynman diagram. The possible values are {\tt 'dashed'}, {\tt
'dotted'}, {\tt 'straight'}, {\tt 'wavy'}, {\tt 'curly'}, {\tt 'scurly'},{\tt
'swavy'} and {\tt 'double'}. The default value is chosen according to the spin and
color representation of the particle.\\
\egfbalign
\textcolor{white}{\caption{\label{tab:particle_attributes}}}
\end{table}

In the previous example we have only instantiated the object representing the
top quark. However, since the top quark is not a self-conjugate particle, we
still need to implement an object representing the top antiquark. We could
proceed in a similar way as in the example above, but the {\tt Particle} class
has a built-in method, denoted {\tt anti()}, instantiating the
antiparticle object directly from the corresponding particle object. In the
example of the top quark, the instruction 
\begin{verbatim}
  t__tilde__ = t.anti()
\end{verbatim}
instantiates a {\tt Particle} object called \verb+t__tilde__+ which is
identical to the object {\tt t} previously defined, but with the attributes 
{\tt name} ({\tt
texname}) and {\tt antiname} ({\tt antitexname}) interchanged. In addition, 
all the quantum numbers, including the electric charge ({\tt charge}) and
the color representation ({\tt color}), are set to opposite values.

%%%%%%%%%%%%%%%%%%%%%%%%%
%%%%%%%%%%%%%%%%%%%%%%%%%
%%%%%%%%%%%%%%%%%%%%%%%%%
%%%%%%%%%%%%%%%%%%%%%%%%%

\subsection{Implementation of the parameters of a model}
\label{sec:parameters.py}
Parameters of a model, like masses, coupling constants, \emph{etc.}, are defined in an UFO model as instances of the {\tt Parameter} class (itself defined in the file {\tt object\textunderscore
library.py}) in the file {\tt parameters.py}. All the
parameters used in a model implementation are either \emph{external} (or
equivalently \emph{independent}) parameters or \emph{internal} (or equivalently
\emph{dependent}) parameters. The user must provide as an input the numerical
value of the external parameters (\eg, $\alpha_s = 0.118$), while the internal
parameters are related to other (external and/or internal) parameters via
algebraic relations (\emph{e.g.}, $g_s = \sqrt{4\pi\alpha_s}$).
Since internal and external parameters belong to the same generic class {\tt
Parameter}, their declaration is very similar. We will give an example for each case separately 
in order to emphasize the main differences and features.
The list of all the possible attributes for the {\tt Parameter} class is
summarized in Table \ref{tab:parameter_attributes}.

We start with external parameters. In the UFO format, the external parameters
are all taken to be real and the {\tt type} attribute of the {\tt Parameter}
class must be set to the value {\tt 'real'}. Therefore, complex numbers will
have to be split into their real and imaginary parts. As an example, the
declaration of the external parameter $\alpha_s$ reads
\begin{verbatim}
  aS = Parameter(name = 'aS',
                 nature = 'external',
                 type = 'real',
                 value = 0.118,
                 texname = '\\alpha_s',
                 lhablock = 'SMINPUTS',
                 lhacode = [ 3 ]
                 )
\end{verbatim}
The attributes of the {\tt Parameter} class are all mandatory and contain the
name of the parameter ({\tt name}), its nature ({\tt nature}) which is external and the value of the
parameter ({\tt value}). Since any external parameter is a \emph{real} number,
the value must be a real floating point number. The last two arguments,
{\tt lhablock} and {\tt lhacode}, refer to the Les Houches-like format for the input
parameters which is followed by the UFO. This is a
generalization to any model of the original Supersymmetry Les Houches Accord
\cite{Skands:2003cj,Allanach:2008qq}. All the model parameters are grouped
into blocks, each line of a block containing a
counter (a sequence of integers) associated with a given parameter name and its
corresponding numerical value. The attribute {\tt lhablock} of the {\tt Parameter} object directly
refers to the name of the block in which the considered parameter is stored,
whilst the attribute {\tt lhacode} is a list of integers referring to  the counter.

An additional function related to the Les Houches format is included in the file
{\tt write\textunderscore param\textunderscore card.py}. The class {\tt ParamCardWriter}
can be called from within another \python\ module by issuing the instruction
\begin{verbatim}
ParamCardWriter('./param_card.dat', qnumbers=True)
\end{verbatim}
 and outputs a parameter file named {\tt param\textunderscore card.dat}
which contains all the external parameters defined in the model, grouped into
blocks and counters according to their {\tt lhablock} and {\tt lhacode}
attributes. The first argument in the function above refers to the location of
the output file, whereas the second argument specifies whether or not the {\tt
QNUMBERS} blocks~\cite{Alwall:2007mw} should be included in the output. In
addition, if the second argument is set to {\tt True}, the full set of masses
and widths, even if they are dependent parameters, are written to file. 
In the example of {\tt aS} presented above, the corresponding
entry in the output file would read
\begin{verbatim}
Block SMINPUTS 
    3 1.18000e-01 # aS  
\end{verbatim}
Let us also note that the file {\tt write\textunderscore param\textunderscore
card.py} can be directly used from the command line by issuing the instruction 
\begin{verbatim}
$> python ./write_param_card.py
\end{verbatim}
As a result, an output file named {\tt param\textunderscore
card.dat} is created and contains the numerical values of all the external
parameters. A snapshot of this parameter file for a more complete model reads
\begin{verbatim}
###################################
## INFORMATION FOR SMINPUTS
###################################
Block SMINPUTS
    1 1.325070e+02 # aEWM1
    2 1.166390e-05 # Gf
    3 1.180000e-01 # aS

###################################
## INFORMATION FOR YUKAWA
###################################
Block YUKAWA
    5 4.200000e+00 # ymb
    6 1.645000e+02 # ymt
   15 1.777000e+00 # ymtau
\end{verbatim}

\begin{table}
\bgfbalign
\multicolumn{2}{c}{\textbf{Table~\ref{tab:parameter_attributes}: Parameter class attributes}}\\
\\
{\tt nature} & A string, either {\tt 'external'} or {\tt 'internal'},
specifying whether a given parameter is considered as a dependent or independent
parameter. \\
{\tt name} &  A string, specifying the name of the parameter. \\
{\tt type} &  A string, either {\tt 'real'} or {\tt 'complex'}, specifying
whether a given parameter is a real or a complex number. We remind that
following the UFO synthax, external parameters must be real numbers.\\
{\tt value}  & For external parameters, this attribute is a single real floating-point number. For internal parameters, it consists of a string representing the
analytic expression relating the considered parameter to other external and/or
internal parameters, following a valid \python\ syntax.\\
{\tt texname} & A \TeX{} string representing the parameter name in \TeX{} format.\\
\multicolumn{2}{l}{\textbf{Attributes specific to external parameters}}\\
{\tt lhablock} & A string containing the name of the block which the parameter
is assigned to, following a Les Houches-like format.\\
{\tt lhacode} & A list of integers giving the position of the considered
parameter inside a given {\tt lhablock}, \ie, the counter associated with the 
parameter, following a Les Houches-like format.\\
\egfbalign
\textcolor{white}{\caption{\label{tab:parameter_attributes}}}
\end{table}

The definition of internal parameters follows the same lines as for the external parameters, with
the only differences that the {\tt lhablock} and {\tt lhacode} attributes are
not available and that the {\tt value} argument now contains an algebraic
expression relating the parameter to other external or internal
parameters. As a simple example, consider the external parameter {\tt aS} ($\alpha_s$) and the internal parameter {\tt G} ($g_s=\sqrt{4 \pi \alpha_s}$). The implementation of {\tt G} reads,
\begin{verbatim}
G = Parameter(name = 'G',
              nature = 'internal',
              type = 'real',
              value = 'cmath.sqrt(4 * cmath.pi * aS)',
              texname = 'G'
              )
\end{verbatim}
Unlike the case of external parameters, the {\tt value} attribute is a string representing a valid algebraic \python\
expression. Moreover, it is mandatory that every internal parameter depends only
on other parameters which have already been declared. Returning to our example,
the external parameter {\tt aS} must hence be defined \emph{before} the internal parameter
{\tt G} inside the file {\tt parameters.py}. Note that masses and widths are considered
to be parameters of the model (either internal or external), and must thus be declared as such in {\tt parameters.py}.

Let us conclude this section by mentioning that most matrix-element generators
have information on the Standard Model input parameters hard-coded. This allows,
among others, for a correct handling of the running of the strong coupling
constant. Therefore, the Standard Model parameters in an UFO model must be
correctly identified, following the same notations and conventions as for the
implementation of a model in \feynrules\ \cite{Christensen:2009jx}.

\subsection{Implementation of the interactions of the model} \label{sec:vertices.py}

The vertices corresponding to the interactions included in a model are defined
in the file {\tt vertices.py} using the {\tt Vertex} class.
Let us consider
a set of $n$ particles $\{\phi_i^{\ell_ia_i}\}$, with spin indices\footnote{The terminology {\it spin indices} refers to both Lorentz and Dirac indices.}
$\{\ell_i\}$ and color indices
$\{a_i\}$. A generic $n$-point vertex coupling these fields can be written as a
tensor in the color $\otimes$ spin space\footnote{The case of non-strongly interacting particles
corresponds to a tensor in color space equal to unity.},
\begin{equation}\label{eq:generic_vertex}
  \begin{cal}V\end{cal}^{a_1\ldots a_n, \ell_1\ldots\ell_n}(p_1,\ldots,p_n) =
    \sum_{i,j}C_i^{a_1\ldots a_n}\,G_{ij}\,L_j^{\ell_1\ldots\ell_n}(p_1,\ldots,p_n)
     \ , 
\end{equation}
where the variables $p_i$ denote the particle momenta, $G_{ij}$ the couplings, and the
quantities $C_i^{a_1\ldots a_n}$ and $L_j^{\ell_1\ldots\ell_n}(p_1,\ldots,p_n)$
denote tensors in color and spin space, respectively. Since several vertices may share the
same spin and/or color tensors, the latter act as a `basis' for the vertices
of the model, the couplings being the `coordinates' in that basis.
As an example, the well-known four-gluon vertex,
\begin{equation}\begin{split}
  ig_s^2&\,f^{a_1a_2b}f^{ba_3a_4}\,\left(\eta^{\mu_1\mu_4}\eta^{\mu_2\mu_3} - \eta^{\mu_1\mu_3}\eta^{\mu_2\mu_4}\right)\\
&+ig_s^2\,f^{a_1a_3b}f^{ba_2a_4}\,\left(\eta^{\mu_1\mu_4}\eta^{\mu_2\mu_3} - \eta^{\mu_1\mu_2}\eta^{\mu_3\mu_4}\right)\\
&+ig_s^2\,f^{a_1a_4b}f^{ba_2a_3}\,\left(\eta^{\mu_1\mu_3}\eta^{\mu_2\mu_4} -
\eta^{\mu_1\mu_2}\eta^{\mu_3\mu_4}\right)\ , 
\end{split}\end{equation}
is written in Eq.\ \eqref{eq:generic_vertex} as 
\begin{equation}\begin{split}\label{eq:4gluon}
\left(f^{a_1a_2b}f^{ba_3a_4},\right. &\left.f^{a_1a_3b}f^{ba_2a_4}, f^{a_1a_4b}f^{ba_2a_3}\right)\\
&\times\left(\begin{array}{ccc}
ig_s^2 & 0 & 0\\
0 & ig_s^2 & 0\\
0 & 0 & ig_s^2
\end{array}\right)\,
\left(\begin{array}{c}
\eta^{\mu_1\mu_4}\eta^{\mu_2\mu_3} - \eta^{\mu_1\mu_3}\eta^{\mu_2\mu_4} \\
\eta^{\mu_1\mu_4}\eta^{\mu_2\mu_3} - \eta^{\mu_1\mu_2}\eta^{\mu_3\mu_4}\\
\eta^{\mu_1\mu_3}\eta^{\mu_2\mu_4} - \eta^{\mu_1\mu_2}\eta^{\mu_3\mu_4}\
\end{array}\right)\ .
\end{split}\end{equation}
The UFO format for vertices mimics exactly this structure. As an example, the
implementation of the vertex in Eq.~\eqref{eq:4gluon} into an UFO model reads
\begin{verbatim}
V_1 = Vertex(name = 'V_1',
              particles = [ P.G, P.G, P.G, P.G ],
              color = [ 'f(1,2,-1)*f(-1,3,4)', 
                        'f(1,3,-1)*f(-1,2,4)', 
                        'f(1,4,-1)*f(-1,2,3)'  ],
              lorentz = [ L.VVVV1, L.VVVV2, L.VVVV3 ],
              couplings = {(0,0):C.GC_1,
                           (1,1):C.GC_1,
                           (2,2):C.GC_1}
              )
\end{verbatim}
\begin{table}
\bgfbalign
\multicolumn{2}{c}{\textbf{Table~\ref{tab:vertex_attributes}: Vertex class
attributes}}\\ \\
{\tt name} &  A string specifying the name tag of the vertex. \\
{\tt particles} &  A list of {\tt Particle} objects containing the set of
particles entering into the vertex. By convention, all particles are considered outgoing. \\
{\tt color}  & A list of strings representing the color tensors associated with
the vertex, written as a polynomial combination of
the elementary tensors given in Table \ref{tab:color_tensors}.\\
{\tt lorentz}  & A list of {\tt Lorentz} objects representing the spin tensors
associated with the vertex.\\
{\tt couplings}  & A list of {\tt Coupling} objects associated with the
decomposition of the vertex in the color $\otimes$ spin space. \\
\egfbalign
\textcolor{white}{\caption{\label{tab:vertex_attributes}}}
\end{table}
The {\tt Vertex} class is probably one of the most important features of the UFO
format, since the vertices associated with a Lagrangian are at the heart of every
implementation of a BSM
model into a matrix-element generator. It requires five arguments, which are
summarized in Table \ref{tab:vertex_attributes}. First, each
vertex is identified by an identification tag, its {\tt name}. Next, the
attribute {\tt particles} contains the list of all {\tt Particle} objects entering the
considered vertex (by convention, all particles are considered outgoing). 
Since these objects are defined in the file
{\tt particles.py}, it is necessary to issue at the beginning of the file {\tt
vertices.py} the command
\begin{verbatim}
import particles as P
\end{verbatim}
and a particle object {\tt G} is now referred to as
{\tt P.G}.
The attributes {\tt color} and {\tt lorentz} contain two lists with the color
and Lorentz tensors associated with the vertex, \ie, the quantities
$C_i^{a_1\ldots 
a_n}$ and $L_j^{\ell_1\ldots\ell_n}(p_1,\ldots,p_n)$ appearing  in
Eq.\ \eqref{eq:generic_vertex}, and are represented inside an UFO module as,
\begin{equation*}\begin{split}
\left(C_0^{a_1a_2a_3a_4}, C_1^{a_1a_2a_3a_4}, C_2^{a_1a_2a_3a_4}\right) &\,
\leftrightarrow  \textrm{{\tt  ['f(1,2,-1) * f(-1,3,4)', ... ]}}\ , \\
\left(L_0^{\mu_1\mu_2\mu_3\mu_4},L_1^{\mu_1\mu_2\mu_3\mu_4},
L_2^{\mu_1\mu_2\mu_3\mu_4}\right) &\,\leftrightarrow  \textrm{{\tt [ L.VVVV1,
L.VVVV2, L.VVVV3 ]}} \ .
\end{split}\end{equation*}
Each color tensor is given as a string representing a polynomial combination of
elementary color tensors, whose arguments are integer numbers referring to
the position of the particle in the list {\tt particle}. If two indices
are contracted, they are represented by a negative integer. The set of all the
elementary color tensors currently included in the UFO format, together with the
corresponding \python\ syntax, is given in Table \ref{tab:color_tensors}.
Using these conventions, the color tensors related to the four gluon vertex are
given by 
\begin{equation*}\begin{split}
f^{a_1a_2b}\ f^{ba_3a_4}  \quad \leftrightarrow \quad &  \textrm{\tt'f(1,2,-1) *
f(-1,3,4)'}\ ,\\
f^{a_3a_2b}\ f^{ba_2a_4}  \quad \leftrightarrow \quad &  \textrm{\tt'f(1,3,-1) *
f(-1,2,4)'}\ ,\\
f^{a_1a_4b}\ f^{ba_2a_3}  \quad \leftrightarrow \quad &  \textrm{\tt'f(1,4,-1) *
f(-1,2,3)'}\ .
\end{split}\end{equation*}
Since the list of color tensors associated with a vertex is a mandatory argument
of the {\tt Vertex} object, we define the trivial color structure associated with
an interaction among non-colored particles as the color tensor \verb+'1'+.

\begin{table}
\bgfbwide
\multicolumn{2}{c}{\textbf{Table~\ref{tab:color_tensors}: Elementary color tensors}}\\
\\
Trivial tensor (for non-colored particles) & {\tt 1} \\
Kronecker delta $\delta^{\bar \jmath_2}{}_{i_1}$ & {\tt Identity(1,2)}\\
Fundamental representation matrices  $\left(T^{a_1}\right)^{\bar\jmath_3}{}_{i_2}$ & {\tt T(1,2,3)} \\
Structure constants $f^{a_1a_2a_3}$ & {\tt f(1,2,3)}\\
Symmetric tensor $d^{a_1a_2a_3}$ & {\tt d(1,2,3)}\\
Fundamental Levi-Civita tensor $\epsilon_{i_1i_2i_3}$ & {\tt Epsilon(1,2,3)}\\
Antifundamental Levi-Civita tensor $\epsilon^{\bar\imath_1\bar\imath_2\bar\imath_3}$ & {\tt EpsilonBar(1,2,3)}\\
Sextet representation matrices  $\left(T_6^{a_1}\right)^{\bar\beta_3}{}_{\alpha_2}$ & {\tt T6(1,2,3)} \\
Sextet Clebsch-Gordan coefficient $(K_6)^{\bar\imath_2\bar\jmath_3}{}_{\alpha_1}$ & {\tt K6(1,2,3)}\\
Antisextet Clebsch-Gordan coefficient $(\bar K_6)^{\bar\alpha_1}{}_{i_2j_3}$ & {\tt K6Bar(1,2,3)}\\
\egfbwide
\textcolor{white}{\caption{\label{tab:color_tensors}}}
\end{table}

Spin structures such as those appearing in the vertex decompositions in color
$\otimes$ spin space are implemented as instances of the class {\tt Lorentz}.
All the structures necessary for the whole model are declared in the {\tt lorentz.py}
file and we must hence issue at the beginning of the file {\tt
vertex.py} the instruction 
\begin{verbatim}
import lorentz as L
\end{verbatim}
Hence, the {\tt Lorentz} objects used in {\tt vertex.py}, declared in the {\tt
lorentz.py} \python\ module, are preceded by the prefix {\tt L}. As
illustrated in the example of the four-gluon vertex, the {\tt
lorentz} attribute of the {\tt Vertex} class contains the list of the relevant
structures. A {\tt Lorentz} object is instantiated as
\begin{verbatim}
FFV1 = Lorentz(name = 'FFV1',
               spins = [ 2, 2, 3 ],
               structure = 'Gamma(3,2,1)')
\end{verbatim} 
All attributes are mandatory. While the attribute {\tt name} is defined in the usual way,
the attribute {\tt spins} contains the list of the values of the spins, written as 
$(2s+1)$, of the particles entering the vertex. The last argument, {\tt 
structure}, gives the analytical formula of the Lorentz structure as a string.
The conventions for the spin indices is similar to the convention for
the color indices: a positive integer $i$ points to the entry $i$ in the list {\tt spins} while negative integers are
contracted indices. By default, all the Lorentz indices are supposed to be upper
indices, and repeated Lorentz indices are contracted using the Minkowski
metric. The list of all objects that can be used to define a Lorentz structure
is given in Table~\ref{tab:lorentz_structure}.

\begin{table}
\bgfbwide
\multicolumn{2}{c}{\textbf{Table~\ref{tab:lorentz_structure}: Elementary Lorentz structures}}\\
\\
Charge conjugation matrix: $C_{i_1 i_2}$& {\tt C(1,2)}\\ 
Epsilon matrix: $\epsilon^{\mu_1 \mu_2 \mu_3\mu_4}$ & {\tt Epsilon(1,2,3,4)}\\
Dirac matrices: $(\gamma^{\mu_1})_{i_2 i_3}$ & {\tt Gamma(1, 2, 3)}\\
Fifth Dirac matrix:  $(\gamma^5)_{i_1 i_2}$ & {\tt Gamma5(1,2)} \\
(Spinorial) Kronecker delta: $\delta_{i_1 i_2}$ &  {\tt Identity(1,2)} \\
Minkowski metric: $\eta_{\mu_1 \mu_2}$ & {\tt  Metric(1,2) }\\
Momentum of the $N^{\text{th}}$ particle: $p^{\mu_1}_N$& {\tt  P(1,N) }\\
Right-handed chiral projector:  $\left(\frac{1+\gamma5}{2}\right)_{i_1 i_2}$ & {\tt  ProjP(1,2) }\\
Left-handed chiral projector $\left(\frac{1-\gamma5}{2}\right)_{i_1 i_2}$ &  {\tt ProjM(1,2) }\\
Sigma matrices: $(\sigma^{\mu_1 \mu_2})_{i_3 i_4}$ &  {\tt Sigma(1,2,3,4)}\\
\egfbwide
\textcolor{white}{\caption{\label{tab:lorentz_structure}}}
\end{table}

For a given vertex, the $G_{ij}$ quantities appearing in Eq.\
\eqref{eq:generic_vertex} are the `coordinates' corresponding to the
decomposition of a
vertex into the color $\otimes$ spin basis.
The {\tt couplings} attribute of the {\tt Vertex} class contains hence a
\python\ dictionary relating the coordinate $(i,j)$ to a {\tt
Coupling} object, declared in the file {\tt couplings.py},
\begin{equation*}
G_{ij}  \quad  \leftrightarrow  \quad \textrm{{\tt(i,j):C.GC\textunderscore 1}}\ ,
\end{equation*}
By convention, only  non-vanishing coordinates $G_{ij}$
are included in this dictionary. Moreover, the  {\tt Coupling} objects
must be imported at the beginning of the {\tt vertices.py} file through
the command 
\begin{verbatim}
import couplings as C
\end{verbatim}
The declaration of the {\tt Coupling} objects in the file
{\tt couplings.py} is similar to the one of internal parameters. Going back to
the example of the four-gluon vertex in Eq.\ \eqref{eq:4gluon}, the coupling \verb+GC_1+ 
is defined by
\begin{verbatim}
GC_1 = Coupling(name = 'GC_1',
                value = 'complex(0,1)*G**2',
                order = {'QCD':2}
                )
\end{verbatim}
The attribute {\tt value} is a string
giving the algebraic expression of the coupling in terms of internal and/or
external parameters. The last
attribute of a {\tt Coupling} object, {\tt order}, is a
\python\ dictionary where the key for each entry is a string and the value
a non-negative integer. In the example above, this means that the four-gluon
vertex is proportional to two powers of the strong coupling. This feature
allows certain matrix-element generators to generate only subclasses of Feynman diagrams at runtime. This subclass is determined by giving an upper limit for a given interaction type, specified by the key in the dictionary {\tt order}. This concept, together with its implementation into the UFO format, is explained in the next section.

\subsection{Controlling various types of couplings in a perturbative
expansion}\label{sec:order}
In this section we discuss how to control the different types of expansion
parameters that might appear in a perturbative expansion. To illustrate this
concept, let us consider the production of a weak boson in association with jets
at a hadron collider, \emph{e.g.}, $p\,p\to Z +  4$ jets. This process is
dominated by QCD production, while diagrams involving off-shell weak boson
exchanges are highly suppressed. In order to speed up the event generation for
this process, it is thus desirable to focus exclusively on the strong
production of the additional four jets, neglecting all Feynman diagrams with
weak boson exchanges. In other
words, we would like to select the subset of all the diagrams contributing to
the process $p\,p\to Z + 4$ jets that involve at most one electroweak vertex,
\emph{i.e.},  at most one power of the electromagnetic coupling constant $e$.

This can be achieved using tags that allow to count
the number of couplings of a given type present in a diagram.
In the previous section, we have introduced the {\tt
order} attribute of the {\tt Coupling} class. As examples, the {\tt order} of
$g_s^2$ was hence defined as \verb+{`QCD', 2}+, whilst the one of $e^2$ reads
\verb+{`QED':2}+. In the case of the generation of the Feynman diagrams
associated to the $p\,p\to Z + 4$ jets process, the coupling order feature
allows to restrict 
the number of couplings of type {\tt QED} to be at most one, neglecting in this
way the electroweak production of  any additional jet\footnote{We stress that coupling
orders are a property of the matrix-element generators, \emph{i.e.}, the
matrix-element generator in question needs to support this feature to
use it.}. For certain models, it can be useful to specify a default behavior for
some types of coupling orders. This can be done using the {\tt CouplingOrder}
class, which we describe in the rest of this section.

Coupling orders are instances of the class {\tt CouplingOrder}
and are instantiated in the file {\tt coupling\textunderscore orders.py}. As a
first simple examples, let us consider the coupling orders {\tt QCD} and {\tt QED},
corresponding to the coupling constants $g_s$ and $e$, respectively. The definitions in {\tt coupling\textunderscore orders.py} read
\begin{verbatim}
QCD = CouplingOrder(name = `QCD',
                    expansion_order = 99,
                    hierarchy = 1
                    )
                                     
QED = CouplingOrder(name = `QED',
                    expansion_order = 99,
                    hierarchy = 2
                    )                                 
\end{verbatim}
The class {\tt CouplingOrder} has two mandatory attributes, apart from the
ubiquitous {\tt name} attribute. First, the attribute {\tt
expansion\textunderscore order} is an integer specifying the maximal
number of couplings of this type that should be included in a given process. The
default value is 99, indicating that any number is allowed. 
The second attribute, {\tt hierarchy}, is an integer that allows one to classify
different types of interactions according to their relative strength. In the
above example, we have {\tt QCD.hierarchy} = 1 and {\tt QED.hierarchy} = 2,
reflecting the fact that $g_s^4$ is of the same order of magnitude as 
$e^2$. The {\tt CouplingOrder} objects then allow certain matrix-element generators to define 
a default behavior for the maximal number of couplings of a given type that can appear in a diagram, based on the upper bound set by {\tt expansion\textunderscore order} and the relative strength among the various couplings.

\section{The \feynrules\ UFO interface}
\label{sec:feynrules}
Even though it is possible to implement a model into the UFO format by hand,
this procedure can be a tedious and error-prone task, because all the vertices
need to be entered one at the time. In order to alleviate this problem, we have
implemented an interface into \feynrules\ that allows one to export a given
model directly in the UFO format. The \feynrules\ model contains, on the one
hand, basic model information (such as the particle content or the parameters of
the model) which is implemented as described in Refs.\
\cite{Christensen:2008py,Christensen:2009jx,Duhr:2011se}.
In particular, a new feature of the FeynRules model files allows to 
specify the hierarchy between the different
types of couplings and the limit up to which they should appear in the perturbative
expansion (see Section~ \ref{sec:order}). 
This is achieved by including the global variables {\tt
M\$InteractionOrderHierarchy} and {\tt
M\$InteractionOrderLimit} directly into the \feynrules\ model file\footnote{{\tt InteractionOrder} is the \feynrules\ equivalent
to the {\tt order} attribute of the UFO {\tt Coupling} object presented in Section
\ref{sec:vertices.py}.}.
Considering the example of the types of couplings QED and QCD presented in Section \ref{sec:order}, the
\feynrules\ model implementation includes then the definition
\begin{verbatim}
  M$InteractionOrderHierarchy = {
    {QCD, 1},
    {QED, 2}
  }
  
  M$InteractionOrderLimit = {
    {QCD, 99},
    {QED, 99}
  }
\end{verbatim}
Note that this new feature is optional for each type of coupling. If a given type if not represented in one of the two lists, the default values assigned will be 1 for the {\tt hierarchy} and 99 for the {\tt expansion\textunderscore order}.

The \feynrules\ UFO interface can be called in exactly the same way as all the
other \feynrules\ interfaces,
\begin{quote}
{\tt WriteUFO[} $\begin{cal}L\end{cal}_1, \begin{cal}L\end{cal}_2, \ldots, options$\,{\tt  ]}
\end{quote}

where $\begin{cal}L\end{cal}_1, \begin{cal}L\end{cal}_2, \ldots$ denote the
Lagrangians of the model, and \emph{options} denotes a set of options supported
by the interface. The interface shares all the options of the function {\tt
FeynmanRules[ ]}, plus some additional options summarized in Table \ref{tab:WriteUFO}.
When this command is issued, \feynrules\ internally calls the function {\tt
FeynmanRules[ ]} to compute all the vertices associated with the Lagrangians $\begin{cal}L\end{cal}_i$. After the
complete list of Feynman rules has been obtained, the vertices are decomposed
into a color $\otimes$ spin basis\footnote{Note that this decomposition might
not be unique.} according to Eq.~(\ref{eq:generic_vertex}), and the different
{\tt Lorentz} and {\tt Coupling} objects are identified. 
At the end of the procedure, all the
information about the model is written to files according to the format
presented in Section~\ref{sec:ufo_format}, and saved in a directory called {\tt
*\textunderscore UFO}, where {\tt *} denotes the name of the model.

Note that there is a crucial difference between the UFO interface and the other existing
interfaces included in the \feynrules\ package. While all other interfaces
select the subset of vertices that are supported by the matrix-element
generators (in general, this subset consists more or less into renormalizable
operators) while rejecting all other vertices, the UFO
interface is completely agnostic of the matrix-element generator, and hence does
not make any assumptions on whether a given generator accepts a certain
vertex structure. The UFO output will hence always contain \emph{all} the vertices
of the model, and it is then up to the matrix-element generator to assure that
only allowed vertices are processed.

\begin{table}
\bgfbalign
\multicolumn{2}{c}{\textbf{Table~\ref{tab:WriteUFO}: Additional options of the function {\tt WriteUFO}}}\\
{\tt Input} & A list of vertices computed previously and to be included into the UFO output.\\
{\tt Output} & A string, the name of the output directory. The default is the value of the \feynrules\ variables {\tt M\$ModelName} with {\tt \textunderscore UFO} appended.\\
{\tt DialogBox} & If {\tt Off}, no dialog boxes open up when running the interface. The default is {\tt On}.\\
\egfbalign
\textcolor{white}{\caption{\label{tab:WriteUFO}}}
\end{table}

\section{The UFO format beyond tree level}\label{sec:nlo}

During the last five years, a lot of progress has been made in the automation of
the computation of next-to-leading order matrix elements, both regarding
the generation of the real corrections with the appropriate subtraction terms
\cite{Gleisberg:2007md,Seymour:2008mu,Hasegawa:2008ae,Frederix:2008hu,Czakon:2009ss,Frederix:2009yq},
and the development of algorithms for calculating loop amplitudes
numerically
\cite{Zanderighi:2008na,Ellis:2009zw,Berger:2009zg,vanHameren:2009dr,Berger:2010zx,Hirschi:2011pa}.

Although currently the focus of the UFO 
is to provide a common input for tree-level Monte Carlo programs,
the format is by no means restricted to tree-level generators only.
Hence, the one-loop matrix-element generator \gosam~\citegolem\
contains an interface to the UFO format, where the information from
the \python{} module described in Section \ref{sec:ufo_format} is translated
into a model definition for the {\sc QGraf}\ package \cite{Nogueira:1991ex}
together with a
{\sc Form}\ \cite{Vermaseren:2000nd} module substituting the expressions from
the Feynman rules.
This setup has been successfully applied to simple one-loop calculations
in the Minimal Supersymmetric Standard Model, where the renormalization can
still be worked out by hand.
For more involved computations, however, one would like to automate not only
the calculation of the matrix elements but also the derivation of the
counterterms associated with a given renormalization procedure. Although
\feynrules{} in its current version does not yet support the calculation of
renormalization constants and counterterms, we propose in this section
a generic prescription for their inclusion in the UFO~format. 

Assuming a multiplicative renormalization prescription, the relation between
bare and renormalized quantities is given by $m_0=Z_m m_r=(1+\delta Z_m)m_r$,
where $m$ represents a generic parameter, and by
$\Psi_0=Z_\Psi^{1/2}\Psi_r=(1+\frac12\delta Z_\Psi)\Psi_r$ for the fields. The
general case of propagator mixing allows the last equation to take a matrix form.
Furthermore, it is assumed that the ultraviolet divergences have been
regularized dimensionally, the renormalization constants being thus expressed as
Laurent series in $\epsilon=(4-D)/2$ where $D$ is the number of space-time
dimensions. Taking into account that the format should not be restricted to one
type of perturbative corrections but should be extendable to any
$\alpha_s^{n_1}\alpha_{EW}^{n_2}$ order of the perturbative expansion, we can make
the ansatz
\begin{equation}\label{eq:deltaZ}
\delta Z_i = \sum_{n_1,n_2=1}^\infty
\frac{\alpha_s^{n_1}\alpha_{EW}^{n_2}}{(2\pi)^{n_1+n_2}}
\sum_{p=-\infty}^\infty
z_{n_1,n_2}^{(p)}\,\epsilon^p,
\end{equation}
where, in general, only a small subset of the coefficients $z_{n_1,n_2}^{(p)}$
is non-zero. To include the renormalization constants associated with a
parameter, we propose to add an attribute to the {\tt Parameter} class
denoted {\tt counterterm}. Taking the example of the strong coupling constant
introduced in Section \ref{sec:parameters.py}, {\tt G}, its definition is
augmented by
\begin{verbatim}
G.counterterm = { (1,0): {-1: '2./3*NF*TF-11./6*CA'} }
\end{verbatim}
in order to include the QCD one-loop effects on the strong coupling constant.
The renormalization constant is represented by a \python\ dictionary where the
keys are the pairs $(n_1, n_2)$ introduced in Eq.\ \eqref{eq:deltaZ} and the
values are the Laurent series in $\epsilon$. The latter are represented by
dictionaries with the powers of $\epsilon$ as keys and strings
representing \python{} expressions as values. Let us note that the symbols {\tt
NF}, {\tt TF} and {\tt CA} which have been introduced must be either replaced by
their proper values or be defined as model parameters.
For models containing more
than two coupling constants, the pairs $(n_1,n_2)$ are replaced by the corresponding
$n$-tuples.
Similarly, wave function renormalization constants\footnote{Mixing of on-shell
particles is assumed to be zero. However, in propagators, mixing is realized through
two-point vertices.} are included in the {\tt counterterm}
attribute which is added to the {\tt Particle} class. It contains the
object $\delta Z_\Psi$, implemented following a structure identical to the one
described for the {\tt Parameter} class. 

In addition to the renormalization constants, one also needs analytical
expressions for the counterterms. In general, they can be described as vertices,
starting from two-point vertices for the propagator counterterms, which are
included in the files {\tt ctvertices.py} and {\tt ctcouplings.py}. Analogously,
the initialization file {\tt
\textunderscore\textunderscore{}init\textunderscore\textunderscore.py} contains,
in addition to the lists described in the
previous section, the lists {\tt all\textunderscore{}ctvertices} and {\tt
all\textunderscore{}ctcouplings}. Similarly to Section \ref{sec:vertices.py},
the file {\tt ctvertices.py} contains all the counterterm vertices represented
by objects of the {\tt Vertex} class, and the related couplings are included in
the file {\tt ctcouplings.py}. These couplings reflect the nature of the
renormalization constants as Laurent expansion in $\epsilon$. 
Using the generic structure for vertices presented in Eq.\
\eqref{eq:generic_vertex}\footnote{The basis in the color $\otimes$ spin space
associated to a counterterm vertex
might be different from the corresponding tree-level one.}, we can write a counterterm
coupling as 
\begin{equation}
G_{ij} = g_{ij}^{(0)}\sum_{n_1,n_2=1}^\infty
\frac{\alpha_s^{n_1}\alpha_{EW}^{n_2}}{(2\pi)^{n_1+n_2}}
\sum_{p=-\infty}^\infty
c_{ij,n_1,n_2}^{(p)}\,\epsilon^p \ .
\end{equation}
This ansatz allows for some freedom with respect to numeric factors
that can be part of either $g_{ij}^{(0)}$ or $c_{ij,n_1,n_2}$. However, the
power of the coupling constants in $g_{ij}^{(0)}$ must correspond to the one
included in the associated tree level vertex. Hence, the counterterm coupling
can be easily declared using the {\tt Coupling} class, 
\begin{verbatim}
GCT_1 = Coupling(name = 'GCT_1',
       value = 'complex(0,1)*G**2',
       counterterm = {(1,0): {-1: G.counterterm}},
       order = {'QCD':2}
   )
\end{verbatim}
The prefactor $g_{ij}^{(0)}$ is stored in the {\tt value} attribute
whereas all relative corrections $c_{ij,n_1,n_2}^{(p)}$
are mapped to the attribute {\tt counterterm}, using the same philosophy
as in the case of the classes {\tt Parameter} and {\tt Particle}.
Finally, the attribute {\tt order} reflects the interaction order of
$g_{ij}^{(0)}$ and does not take into account the additional powers of coupling
constants coming from the sum over $n_1$ and $n_2$.

The amendments described in this section transmit all information necessary for
an efficient computation of ultraviolet counterterms by a matrix-element
generator. Furthermore, the same approach could be used in order to include
other counterterm-like objects, such as the rational $R_2$
terms~\cite{Ossola:2008xq,Draggiotis:2009yb,Garzelli:2009is,Garzelli:2010qm} in
the OPP approach~\cite{Ossola:2006us,Ossola:2007bb}.

\section{An example}
\label{sec:example}

An UFO model contains the full set of vertices of a model, \emph{i.e.}, all the
Lorentz and color structures appearing in all the vertices together with their
coefficients. Consequently, it is also suited for models with Lorentz structures
that are not SM-like, a characteristic shared by all models with
higher-dimensional operators. In the following, we illustrate the UFO format on the
example of the Strongly Interacting Light Higgs (SILH)
model~\cite{Giudice:2007fh}. The SILH model is an effective theory describing
the interactions of the Higgs boson considering it as the Goldstone boson linked
to a new strongly interacting sector. Since it is already implemented in
\feynrules\ \cite{Christensen:2009jx}, this model can easily be exported to the
Monte Carlo tools via the corresponding \feynrules\ interfaces.

The particle content of the SILH model is the same as in the SM. The
particularities of the model come solely from the new interactions induced by
dimension-six operators involving SM fields. In this short example, we focus on
the decay of the Higgs boson $H$ into two $W$-bosons. The non-SM part of the SILH
Lagrangian affecting this decay rate reads
\begin{eqnarray}
{\cal L}_{\rm SILH}^{HWW} &=& \frac{c_H}{2f^2}\partial^\mu \left( H^\dagger H \right) \partial_\mu \left( H^\dagger H \right) 
+\frac{ic_Wg}{2g_\rho^2 f}\left( H^\dagger  \sigma^i \overleftrightarrow {D^\mu} H \right )( D^\nu  W_{\mu \nu})^i\nonumber\\
&&+\frac{ic_{HW} g}{16\pi^2f^2}(D^\mu H)^\dagger \sigma^i(D^\nu H)W_{\mu \nu}^i
\label{lsilh}
\end{eqnarray}
where $f$ is the suppression scale for the new operators, $g$ and $g_\rho$ are
the coupling constants of the weak and the new strong interaction, respectively,
and $c_H$, $c_W$ and $c_{HW}$ are free coefficients. In the expression above, we
have introduced the covariant derivative $D_\mu$ (taken in the
appropriate representation), the $W$-boson field
strength tensor $W_{\mu\nu}$, and the Pauli matrices $\sigma^i$. The effective
Lagrangian has been obtained after an expansion in $1/f$ up to
$\begin{cal}O\end{cal}(1/f^2)$. Hence, the $HWW$ vertex reads now
\begin{eqnarray}
&i g_w^2 \left[\frac{v}{2}\left(1-c_H\frac{\xi}{2}\right) \eta_{\mu_2,\mu_3} + c_{HW} \xi \frac{p_1^{\mu _2} p_2^{\mu _3}+p_1^{\mu _3} p_3^{\mu _2}-\left(p_1.p_2+p_1.p_3\right) \eta _{\mu _2,\mu _3}}{32 \pi ^2 v} \right.&\nonumber\\
&\left.+ c_W \xi \frac{\left(p_2.p_2+p_3.p_3\right) \eta _{\mu _2,\mu
_3}-p_2^{\mu _2} p_2^{\mu _3}-p_3^{\mu _2}  p_3^{\mu _3}}{2 v g_{\rho }^2}
\right]&\label{hww} \ ,
\end{eqnarray}
where $\xi=\frac{v^2}{f^2}$, $v$ being the vacuum expectation value of the
neutral component of the Higgs doublet, and $p_i$ are the momenta of the interacting
particles.

After an UFO implementation of the model via the corresponding
\feynrules\ interface has been obtained,
this vertex appears in \verb+vertices.py+ as
\begin{verbatim}
V_22 = Vertex(name = 'V_22',
              particles = [ P.W__minus__, P.W__plus__, P.H ],
              color = [ '1' ],
              lorentz = [ L.VVS1, L.VVS5, L.VVS8 ],
              couplings = {(0,1):C.GC_56,(0,2):C.GC_59,
                           (0,0):[ C.GC_30, C.GC_68 ]}).
\end{verbatim}
The color tensor is trivial since all particles are color singlets. On the contrary,
the spin structure is more complicated because of the non-trivial
tree-level Lorentz structures of Eq.\eqref{hww}. As an example, the 
\verb+VVS8+ spin tensor is defined in the file {\tt lorentz.py} as 
\begin{verbatim}
VVS8 = Lorentz(name = 'VVS8',
               spins = [ 3, 3, 1 ],
               structure = 'P(1,3)*P(2,1) + P(1,2)*P(2,3)
                            - P(-1,1)*P(-1,3)*Metric(1,2) 
                            - P(-1,2)*P(-1,3)*Metric(1,2)')
\end{verbatim}
The product of \verb+VVS8+ and \verb+GC_59+ corresponds to the second term of Eq.\ \eqref{hww}. The coupling order of the coupling is given by \verb+NP=1+, indicating that it contains one power of $\xi$. It is important to note that \verb+VVS1+, the Lorentz structure of the first term in Eq.\eqref{hww} is associated with two coefficients, split according to their coupling order. In particular, \verb+GC_30+ is the SM part and corresponds to the order \verb+QED=1+, while \verb+GC_68+ is the new physics contribution proportional to $c_H$ and of order \verb+NP=1+. Interferences between SM and new physics operators can be extracted from the interaction order of the vertex. Indeed, due to our choice for the $c_i$ coefficients, the new physics pieces of the vertex have an interaction order equal to \verb+NP=1+ and \verb+QED=0+. Therefore, the interference is obtained by computing the difference between all the contributions ({\tt NP=1 QED=1}) and the pure SM (\verb+NP=0 QED=1+) and SILH (\verb+NP=1 QED=0+) ones.

The Lagrangian of Eq.\ \eqref{lsilh} is truncated at order ${\cal O}(1/f^2)$, or
equivalently at order ${\cal O}(\xi)$.
Computation of matrix elements at higher order in $\xi$ would hence
not be reliable without adding the corresponding terms in the expansion of the
Lagrangian. For instance, the production of a Higgs-boson pair by weak boson
fusion involves a diagram containing two vertices with order \verb+NP=1+, as
presented in Fig.~\ref{fig:wwhh}, but also an additional diagram related to the
expansion of the Lagrangian at order ${\cal O}(\xi^2)$, which is absent from our
SILH model implementation. To prevent the user from such issues, the model
builder should warn him that ${\cal O}(\xi^n)$ amplitudes, with $n\geq 2$,
cannot be in general computed using the implemented Lagrangian, and that the
\verb+NP+ order should be at most equal to one. This restriction is easily
included in the UFO model through to the {\tt expansion\textunderscore order}
argument of the {\tt CouplingOrder} object
\begin{verbatim}
NP = CouplingOrder(name = `NP',
                   expansion_order = 1,
                   hierarchy = 2
                   )
\end{verbatim}
The {\tt hierarchy} argument being set to 2 ensures that the new physics
contributions are not removed by default in the weak processes.

\begin{figure}[!t]\begin{center}
\begin{fmffile}{weakfusion}
\begin{fmfgraph}(120,75)
\fmfleft{g1,g2}
\fmfright{h1,h2}
\fmf{wiggly}{g1,v1,v2,g2}
\fmf{dashes}{v1,h1}
\fmf{dashes}{v2,h2}
\fmffreeze
\fmfv{decor.shape=square,decor.filled=shaded}{v1}
\fmfv{decor.shape=square,decor.filled=shaded}{v2}
\end{fmfgraph}
\end{fmffile}
\caption{\label{fig:wwhh}Feynman diagram contributing to $W^+W^-\to H\,H$ at
${\cal O}(\xi^2)$. The shaded boxes represents the ${\cal O}(\xi)$ part of the
$HWW$ vertex in Eq.\ \eqref{hww}.}
\end{center}\end{figure}

\begin{figure}[!t]
	\centering
		\includegraphics[width=0.90\textwidth]{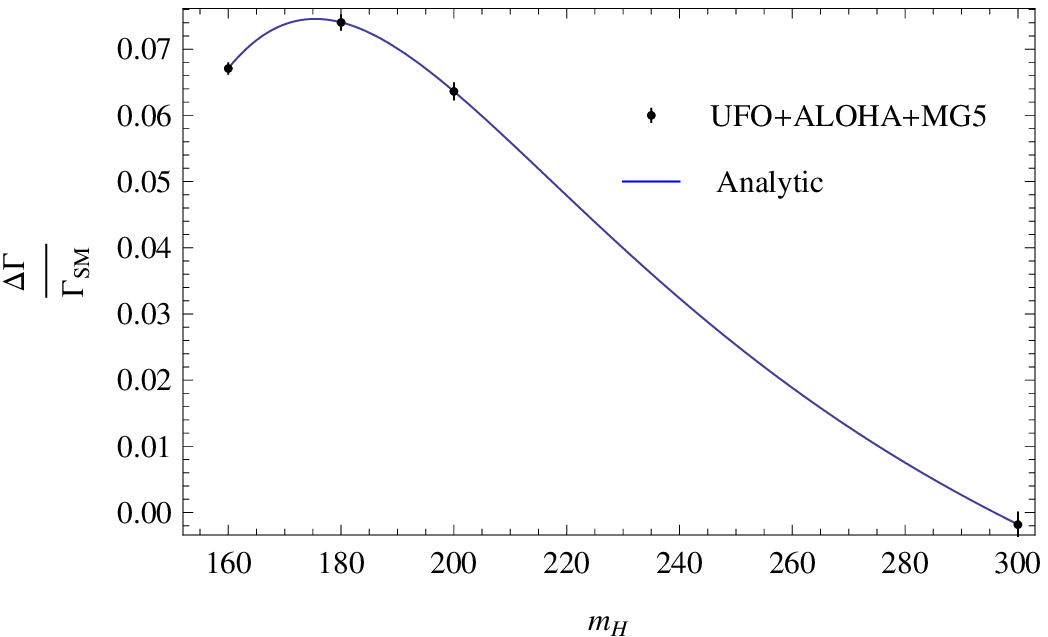}
	\caption{The relative correction
$\frac{\Delta\Gamma}{\Gamma_{SM}}\equiv\frac{\Gamma_{SILH}-\Gamma_{SM}}{\Gamma_{SM}}$
to the decay width of the Higgs boson into two W bosons in the SILH model for
$g_\rho=1$, $f=1$ TeV, $\xi=0.060623$, $c_H=4$, $c_W=2$ and $c_{HW}=800$.
Only the interference terms between SM and diagrams involving the new operators are taken into account in
$\Gamma_{SILH}$ (see Eq.\ (6.6) of Ref.~\cite{Christensen:2009jx} and references
therein).}
	\label{fig:comparison}
\end{figure}

We choose to validate our UFO model by using the computation tools {\sc
MadGraph} version 5~\cite{Alwall:2011uj}, which thanks to the {\sc Aloha}
module~\cite{aloha}, allows for a
full support of the higher dimensional operators. We have computed the
Higgs partial decay width into a $W$-boson pair, using a very large value for
$c_{HW}$ in order to render the associated new physics contribution dominant and to
subsequently test the treatment of the higher-dimensional operators by {\sc MadGraph}.
Moreover, this contribution is not proportional to the SM result, contrary to
the others. In Fig.~\ref{fig:comparison}, we confront the results to hand-made
analytical calculations and found perfect agreement.

\section{Conclusion}\label{sec:conclusions}
In this paper, we have presented a new model format for matrix-element
generators, the Universal \feynrules\ Output (UFO) format. While most of
the present generators have implicit assumptions on the color and/or Lorentz
structures appearing in the different interaction
vertices of a given model, the UFO format has been designed to go
beyond these constraints, by being agnostic of any, even unforeseen,
restrictions. 
%The key feature is flexibility and modularity
%through the use of \python\ classes and objects to represent particles,
%parameters and vertices.
%% REVISED
Indeed, unlike the more traditional table-based model formats (as used by many Monte
Carlo codes), the UFO represents all the information
about a model terms of abstract \python\ classes that can accommodate any (reasonable) 
particle physics model. As an example, despite the fact that so far only color singlet, triplet, sextet
and octet particles have been implemented into the UFO format, the extension to more exotic representations
of the QCD gauge group is in principle straightforward, without requiring any change to the UFO format itself.
A similar change would be very hard to perform in some of the existing table-based model formats.
%which can then be easily exported to any program, and
%even possibly converted to model files to be parsed (as required by many Monte
%Carlo codes). 
%
Finally, we emphasize that the format gives a full support to Les Houches accord
conventions for model parameters and we also illustrate how it could be extented
in the context of the next-to-leading order tools, including, \eg, counterterms
and the so-called $R_2$ terms. Presently, the UFO format is already used by the
\madgraph\ version 5 and \gosam\ generators and will be used in a near future by \herwig++.

\section*{Acknowledgments}
The authors are grateful to Priscila de Aquino, Neil Christensen,
Will Link and to the whole MG5 development team
for useful and constructive discussions.
ClD and BF are grateful to the CP3 Louvain for the hospitality
at various stages during this project.
CeD is a Research Fellow of the
`Fonds National de la Recherche Scientifique'~(FNRS), Belgium.
OM is `Chercheur scientifique logistique postdoctoral F.R.S-FNRS`, Belgium.
TR is supported by the Humboldt Foundation,
in the framework of the Sofja Kovaleskaja Award Project
``Advanced Mathematical Methods for Particle Physics'',
endowed by the German Federal Ministery of Education and Research.
This work was partially supported by the Theory-LHC France Initiative, by the
Research Executive Agency (REA) of the European Union under the Grant Agreement
number PITN-GA-2010-264564 (LHCPhenoNet), by the Belgian Federal Office for
Scientific, Technical  and Cultural Affairs through the Interuniversity
Attraction Poles Program - Belgium Science Policy P6/11-P and by the ISN
MadGraph convention 4.4511.10.

\bibliography{FRBib}

\end{document}